\documentclass[fleqn,usenatbib]{mnras}

\usepackage{newtxtext,newtxmath}

\usepackage[T1]{fontenc}
\usepackage{booktabs}

\DeclareRobustCommand{\VAN}[3]{#2}
\let\VANthebibliography\thebibliography
\def\thebibliography{\DeclareRobustCommand{\VAN}[3]{##3}\VANthebibliography}

\usepackage{graphicx}	
\usepackage{amsmath}	

\usepackage{amssymb}	
\usepackage{booktabs}

\title[WDs collisions in AGN disc]{Electromagnetic signatures of white dwarf collisions in AGN discs}

\author[Shu-Rui Zhang et al.]{
Shu-Rui Zhang,$^{1,2}$\thanks{E-mail: zhangsr@mail.ustc.edu.cn (SRZ)}
Yan Luo,$^{1,2}$
Xiao-Jun Wu,$^{1,2}$
Jian-Min Wang, $^{3}$
Luis C. Ho, $^{4,5}$
Ye-Fei Yuan$^{1,2}$\thanks{E-mail: yfyuan@ustc.edu.cn (YFY)}
\\
$^{1}$School of Astronomy and Space Science, University of Science and Technology of China, Hefei 230026, China\\
$^{2}$CAS Key Laboratory for Research in Galaxies and Cosmology, Department of Astronomy, University of Science and Technology of China, Hefei 230026, China\\
$^{3}$Key Laboratory for Particle Astrophysics, Institute of High Energy Physics, Chinese Academy of Sciences, 19B Yuquan Road, Beijing 100049, China\\
$^{4}$Kavli Institute for Astronomy and Astrophysics, Peking University, Beijing 100871, China\\
$^{5}$Department of Astronomy, School of Physics, Peking University, Beijing 100871, China
}

\date{Accepted 2023 June 15. Received 2023 June 15; in original form 2022 September 30}

\pubyear{2023}

\begin{document}
\label{firstpage}
\pagerange{\pageref{firstpage}--\pageref{lastpage}}
\maketitle

\begin{abstract}
In the inner region of the disc of an active galactic nucleus (AGN), the collision of two white dwarfs (WDs) through Jacobi capture might be inevitable, leading to a Type Ia supernova (SN Ia) explosion. This transient event, influenced by the disc gas and the gravity of the supermassive black hole (SMBH), exhibits distinct characteristics compared to normal SNe Ia. 
The energy of the explosion is mainly stored in the ejecta in the form of kinetic energy. Typically, the ejecta is not effectively decelerated by the AGN disc and rushes rapidly out of the AGN disc. However, under the influence of the SMBH, most of the ejecta falls back toward the AGN disc. As the fallback ejecta becomes more dispersed, it interacts with the disc gas, converting its kinetic energy into thermal energy. This results in a high-energy transient characterized by a rapid initial rise followed by a decay with $L\propto t^{-2.8}$. The time-scale of the transient ranges from hours to weeks, depending on the mass of the SMBH. This process generates high-energy radiation spanning from hard X-rays to the soft $\gamma$ range. Additionally, the subsequent damage to the disc may result in changing-look AGNs. Moreover, the falling back of SNe Ia ejecta onto the AGN disc significantly increases the metallicity of the AGN and can even generate heavy elements within the AGN discs.
\end{abstract}




\section{Introduction}
Compact objects, including white dwarfs (WDs), are believed to exist in the accretion discs of active galactic nuclei \citep[AGN:][]{1983ApJ...273...99O,2012MNRAS.425..460M,mckernan_black_2020,2021ApJ...914L..19Z}, with a predicted high rate of WD-WD mergers  \citep{mckernan_black_2020}. However, the specific mechanisms governing the interaction among WDs in AGN discs and the potential observational phenomena remain unclear. From an observational perspective, phenomena such as AGN variations \citep[e.g.][]{2019BSRSL..88..132G}, changing-look AGNs \citep[e.g.][]{ricci_destruction_2020, 2019ApJ...883...94T}, and the super-solar metallicity of quasars \citep{2003ApJ...596...72W,2006A&A...447..863N,2013ApJ...763...58S,2014MNRAS.438.2828D,2019A&ARv..27....3M} still lack complete understanding. Longstanding issues include the mechanisms by which the extreme environments of AGNs supply and transfer energy, as well as the production of heavy elements in these surroundings.

Numerous physical processes involving compact objects have been proposed within AGN discs \citep[e.g.,][]{1999ApJ...521..502C, 2021ApJ...911L..14W, 2021ApJ...916L..17W, tagawa_formation_2020}. Bound stars around supermassive black hole (SMBHs) can originate from nuclear star clusters \citep{tagawa_formation_2020}. The SMBH can capture stars through either tidal breakup of binary systems (for SMBHs with masses $M_{\rm SMBH}$\,$\sim$\,$10^6M_\odot$) or successive dissipative encounters with the disc \citep[for $M_{\rm SMBH}$\,$\sim$\,$10^8M_\odot$][]{syer_star-disc_1991}. These bound stars (or compact objects formed through evolution) interact continuously with the AGN disc, facilitating the exchange of angular momentum and energy between the compact objects and the disc \citep{1983ApJ...273...99O}. Consequently, the orbits of compact objects are confined to the disc \citep{2020MNRAS.499.2608F, yang_agn_2019, 2017ApJ...835..165B, Tanaka_2002}. Furthermore, stars can form in the outer region of the AGN disc ($\gtrsim10^4r_{\rm g}$, where $r_{\rm g}$ is defined as $GM_{\rm SMBH}/c^2$, and $c$ is the speed of light) due to gravitational instability and subsequently evolve into compact objects \citep{2017MNRAS.464..946S, 2020MNRAS.493.3732D, 2022arXiv220510382D}. The orbits of these compact objects within the AGN disc are influenced by the gravity of the SMBH and the presence of disc gas, causing the objects to migrate inward towards migration traps \citep{2016ApJ...819L..17B} or regions with varying aspect ratios \citep{2014MNRAS.441..900M}. Consequently, various processes, including accumulation, accretion growth, collision, and scattering of compact objects, occur within these regions \citep[e.g.,][]{tagawa_formation_2020, 2019PhRvL.123r1101Y, 2019ApJ...878...85S, 2020ApJ...890L..20G, 2022MNRAS.513.3577V}. Notably, \citet{2022Natur.603..237S} conducted a study on the three-body interactions of black holes (BHs) within AGN discs, revealing that AGN discs can lead to binary mergers with high eccentricities. Additionally, \citet{2021ApJ...916L..17W, 2022arXiv221206133R, 2021arXiv210312088K} proposed that compact objects within AGN discs can capture one another, forming binaries and generating multi-band electromagnetic and gravitational wave (GW) radiation. Further research by \citet{2022ApJ...934..154L} and \citet{2023MNRAS.518.5653B} demonstrated that Jacobi capture in the disc can result in the formation of eccentric BH binaries, which subsequently merge within a few binary orbits.

The evolution of WDs and their binaries in AGN discs is expected to differ from that of BHs. This difference can be attributed, at least in part, to the larger radii of WDs. Furthermore, AGN discs are anticipated to contain a higher number of WDs compared to BHs, as the number density of WDs is higher \citep{2001MNRAS.322..231K}, although the mass function of stars in AGN discs tends to be "top heavy" \citep{2022arXiv220510382D}. Moreover, WDs in AGN discs do not easily reach the Chandrasekhar limit through accretion \citep{2021ApJ...923..173P}. Therefore, it is expected that two-body or multi-body interactions involving WDs would occur within the AGN disc. Specifically, Jacobi capture is a mechanism that allows two WDs to collide directly instead of forming a WD binary \citep{2023MNRAS.tmp.2113L}. Collisions between two WDs are considered as one of the possible pathways for the formation of SNe Ia \citep{rosswog_collisions_2009, 2009MNRAS.399L.156R, hawley_zero_2012}. Although SNe Ia have been utilized as standard candles for measuring the accelerated expansion of the universe \citep{1998AJ....116.1009R, 1998ApJ...507...46S, 1999ApJ...517..565P}, their progenitor systems \citep[e.g.,][]{1973ApJ...186.1007W, 1960ApJ...132..565H, 1984ApJ...277..355W, 1980ApJ...242..749T, rosswog_collisions_2009, 2009MNRAS.399L.156R, 2012NewAR..56..122W} and explosion mechanisms are still subject to controversy \citep[e.g.,][]{2013FrPhy...8..116H, 2014ARA&A..52..107M, 2022MNRAS.514.4078M}. Theoretical models, simulations \citep[e.g.,][]{Mazzali_2001, 2006AIPC..847..406I, rosswog_collisions_2009, 2009MNRAS.399L.156R}, and observations \citep{Hayden_2010} of SNe Ia suggest that only a small fraction of the explosion energy is converted into neutrinos and electromagnetic radiation, while the majority is in the form of kinetic energy. The environment of the AGN disc, characterized by the gravity of the SMBH and high-density gas, can effectively harness the kinetic energy of the ejecta. The disc gas efficiently converts the kinetic energy into electromagnetic radiation, which serves as a potential power source for electromagnetic transient. Additionally, the initial explosion and fallback ejecta may cause disturbances or disruptions in the disc, potentially resulting in observable phenomena.

To date, research on WD explosions in AGN discs has only focused on the interactions between the supernova explosion and the gas of AGN discs during the early phase of evolution. \citet{1995MNRAS.276..597R,moranchel-basurto_supernova_2021} have studied the effects of supernova explosions in the outer region of an accretion disc ($\sim10^4r_{\rm g}$), around a relative massive SMBH ($M_{\rm SMBH}$\,$\gtrsim$\,$10^8M_\odot$). They found that the explosions do not disrupt the disc, but can induce substantial angular momentum redistribution. \citet{perna_electromagnetic_2021} have studied relativistic explosions in AGN discs and extended the standard internal/external shock model \citep{1994ApJ...430L..93R,1999PhR...314..575P,1997ApJ...476..232M} to the AGN disc environment. \citet{2021ApJ...914L..19Z} studied thermonuclear explosions and accretion-induced collapse of WDs in AGN discs, but \citet{2021ApJ...923..173P} shows that WD should not be able to reach the Chandrasekhar limit through accretion. The AGN disc considered is thick and the explosive ejecta can slow down effectively, but the shock wave can explode on the surface of the AGN disc. These processes are verified by numerical simulations that assume that the explosion is located in the region $10^3\sim10^4r_{\rm g}$ \citep{2021MNRAS.507..156G}. 
The previous studies either made assumptions about explosions occurring within a specific radius range or focused on analyzing thick discs. However, they did not examine the potential consequences of failing to dissipate the immense kinetic energy generated by the explosion fully.
Actually, the location of migration traps is usually $\lesssim10^3r_{\rm g}$ if it exists \citep{2016ApJ...819L..17B,2022arXiv220510382D}, and the innermost migration trap can even reach the innermost circular orbit \citep{2021MNRAS.505.1324P}. In addition to a migration trap, an overdensity of compact objects within the disc can also arise in regions where there are changes in the aspect ratio or migration time-scale \citep{2014MNRAS.441..900M}. According to \citet{2012MNRAS.425..460M}, the migration timescale of $1-M_\odot$ compact stars also increases inwards at $\lesssim10^3 r_{\rm g}$. At this location, the gas of the AGN disc is not thick enough, and the explosion ejecta is usually not decelerated effectively. Thus, the ejecta can rush out of the AGN disc. Under the dominance of SMBH gravity, part of the ejecta can fall back on to the AGN disc. As the dispersed ejecta falls towards the AGN disc, 
its interaction with the disc results in the conversion of kinetic energy, leading to the production of high-temperature plasma on the surface of the disc. 

In this work, we investigate the collision of two WDs leading to a SN Ia explosion through Jacobi capture within the inner region of the AGN disc, and the following observable consequences arising after the interaction between the fallback ejecta of the explosion and the AGN disc. Our results show the potential connection between this event and various phenomena observed in AGNs, including AGN variations, changing-look AGNs, and the presence of supersolar metallicity in AGNs.
The introduction of our model is presented in Section ~\ref{sec:2}. The light curves generated from the fallback ejecta hitting the AGN disc are shown in Section ~\ref{sec:3}, and the spectral energy distribution (SED) is provided in Section ~\ref{sec:4}, demonstrating the potential for high-energy radiation generation. The main results, model uncertainties and potential implications are discussed in Section ~\ref{sec:5}. Lastly, the main conclusions are summarized in Section ~\ref{sec:6}.

\section{Theoretical model}
\label{sec:2}
In this section, we provide a comprehensive description of the explosion model resulting from the collision of two WDs within an AGN disc. 
Although the primary objective of this model is to address the explosion caused by the collision of two WDs within an AGN disc, it can be applied to other explosive events occurring within AGN discs, which also involve the fallback of ejecta.

\begin{figure*}
	\includegraphics[width=2\columnwidth]{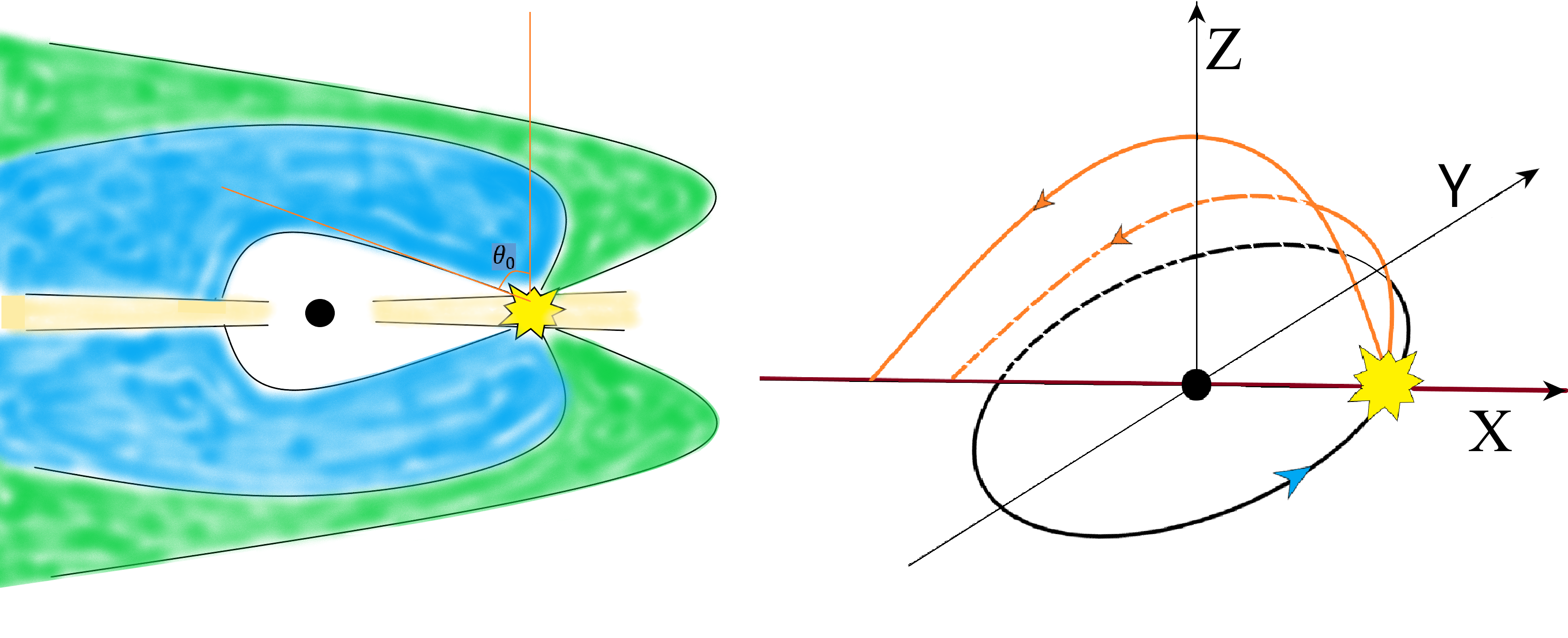}
	\caption{Schematic illustration of WD-WD collision in an AGN disc. \textbf{Left panel}: The collision between two WDs leads to a SN Ia explosion. Subsequently, the ejected material escapes rapidly from the AGN disc. Due to the gravitational pull of the SMBH, a significant portion of the ejecta falls back onto the AGN disc, while some parts manage to escape the system. In our model, we assume that the ejecta with launch angles $\theta<\theta_0=89^{\circ}$ can successfully escape the disc. The unbound material is depicted in green, whereas the bound material is shown in blue, representing the material that can fall back to the AGN disc. \textbf{Right panel}: Schematic representation of the coordinate system employed in the description. The black ellipse in the coordinate system corresponds to the trajectory of the center of mass of the two WDs before the collision. The X-axis is defined as the line along the collision point and the SMBH. The trajectory of the ejecta following the explosion is represented by the orange line. The points where the ejected material can fall back onto the AGN disc are positioned along the X-axis. }
	\label{fig:katong12}
\end{figure*}

\subsection{WD-WD close encounters in AGN discs}
\label{sec:2.1} 

Our discussion here is based on the model of \citet{2022ApJ...934..154L} and \citet{2023MNRAS.518.5653B}, who discuss the close encounter between two stellar massive BHs in an AGN disc. One can also find recent modeling work including gas effects \citep{2022arXiv221206133R}. In the AGN disc, the rate of the close encounter between stellar massive BHs will be enhanced in regions with migration traps or varying aspect ratios, but the relative kinetic energy of two encountering BHs is usually too large to form bound binary BHs. In order to form a bound binary BH, a very close encounter is required, so that the gravitational wave radiation is strong enough to carry away the excess relative kinetic energy. 
Similar to the qualitative analysis in \citet{2022ApJ...934..154L}, we examine the close encounter of two WDs in an AGN disc. To ensure comprehensiveness, we provide only the necessary analysis; detailed information can be found in \citet{2023MNRAS.tmp.2113L}.

Let us consider a SMBH mass of $M_{\rm SMBH}$\,$\sim$\,$10^7M_\odot$ and two identical WDs that are almost coplanar in a circular orbit in an AGN disc. The semi-major axes of the two WDs around the SMBH are $a_1$\,$\approx$\,$a_2\sim100r_{\rm g}$\,$\approx$\,$1.5\times10^9\rm\ km$. Suppose that the masses and radii of the two WDs are $m_1=m_2=1.0M_\odot$ and $R_{\rm WD}$\,$\sim$\,$10^4 \rm\ km$, respectively. Then the mutual Hill radius between the two WDs is give by
\begin{equation}
    R_{\rm H}\equiv\dfrac{a_1+a_2}{2}\left(\frac{m_1+m_2}{3M_{\rm SMBH}}\right)^{\frac{1}{3}}\approx4\times10^{-3}a_1\approx6\times10^6\rm\ km.
	\label{eq:1}
\end{equation}
If the orbital separation of the two WDs is less than $2\sqrt{3}R_{\rm H}$ \citep{1993Icar..106..247G}, the orbits are dynamically unstable. In order to have two unstable WDs captured into a bound binary, GW emission must be dominant, which means that the energy loss by GW emission should follow \citep{2022ApJ...934..154L}
\begin{equation}
	\bigtriangleup\,E_{\rm GW}\gtrsim\varsigma\dfrac{Gm_1m_2}{R_{\rm H}},
	\label{eq:2}
\end{equation}
with $\varsigma$ of order unity. Setting the distance at close encounter as $R_{\rm cap}$,
\begin{equation}
	\dfrac{R_{\rm cap}}{R_{\rm H}}\approx10^{-4}.
	\label{eq:2r}
\end{equation}
Thus, $R_{\rm cap}\approx600\rm\ km$. It should be noted that $R_{\rm cap}\approx600\rm\ km$ is much smaller than a typical WD radius of $R_{\rm WD}$\,$\sim$\,$10^4 \rm\ km$. This indicates that the two WDs collide before a bound binary is formed. In a WD, $mV$ is a constant, where $m$ is the mass and $V$ is the volume of the WD. Namely, the larger the radius of a WD, the easier it is for two WDs to collide instead of forming a binary. Thus, taking the above typical values is reasonable.

The collision of two WDs may produce SNe Ia \citep{rosswog_collisions_2009,2009MNRAS.399L.156R,hawley_zero_2012}. It is worth noting again that WDs in AGN disc are hard to grow to the Chandrasekhar limit through accretion, because WDs can be spun up more efficiently to reach the shedding limit \citep{2021ApJ...923..173P}. This implies that SNe Ia in AGN disc may form mainly by collisions of two WDs rather than through the single-degenerate or usual binary merger channels.

\subsection{The supernova ejecta rushes out of AGN discs}
\label{sec:2.2}
Usually, the total energy from a SN Ia explosion is $\sim1.5\times10^{51}\rm\ erg$, of which the amount taken away by neutrinos is $\sim10^{50}\rm\ erg$ and that by electromagnetic radiation is $\sim10^{49}\rm\ erg$ \citep{Mazzali_2001,2006AIPC..847..406I,Hayden_2010}. Consequently, the majority of the total energy is converted into kinetic energy of the ejecta, leaving only a small fraction of the energy taken away by neutrinos and electromagnetic radiation. Several studies \citep{rosswog_collisions_2009,2009MNRAS.399L.156R,hawley_zero_2012} have simulated SNe Ia resulting from the collision of two WDs. The nuclear detonation triggered by the collision leads to a homologous explosion with velocities ranging from $1.3\times 10^4$ to $1.6\times 10^4\rm\ km\,s^{-1}$, corresponding to a kinetic energy of approximately $10^{51}\rm\ erg$. The resulting electromagnetic radiation is similar to that observed in normal SNe Ia.

By taking into account the collision of two WDs within the gaseous environment of an AGN disc and the subsequent supernova explosion, it becomes feasible to calculate the mass that is swept by the ejected material within the AGN disc. We choose the standard disc model of \citet{kato1998black} to describe the AGN disc. This disc model, which covers our regions of interest ($\lesssim10^3r_{\rm g}$), is akin to the one described in \citet{2003MNRAS.341..501S}. The parameters chosen in our computations include an accretion rate of $\dot{M}=0.1\dot{M}_{\rm Edd}$ and a viscosity of $\alpha=0.1$, where $\dot{M}_{\rm Edd}$ represents the Eddington rate. The scale height and density in the models are kept as continuous. Table ~\ref{tab:table1} displays the disc mass swept by the ejecta for the different central SMBH masses and explosion radii. It can be seen that, in our intriguing cases of explosion in AGN discs (i.e.  $M_{\rm SMBH}$\,$\lesssim$\,$10^8M_\odot$, explosion radius $R_{\rm ej}\lesssim10^3r_{\rm g}$), the swept mass is much less than the mass of the explosive ejecta ($\sim$\,$1M_\odot$). It is  noted that in the other AGN disc models \citep[e.g.][]{2005ApJ...630..167T}  the amounts of the swept-up disc gas is even lower.

Apparently, the AGN disc gas is too thin (i.e. the value of $\rho\times H^3$ is too small, where $\rho$ is the density in the disc midplane and $H$ is the disc scale height, but the disc is still optically thick) for ejecta to slow down effectively. Locally, the ejecta inevitably rushes out of the AGN disc at nearly the initial explosion velocity. At a distance of $\sim100r_{\rm g}$, the escape velocity is $v_{\rm esc}\sim3\sqrt{2}\times 10^4 \rm\ km\,s^{-1}$. The ejecta that rushes out of the AGN disc is subject to the gravitational influence of the SMBH, resulting in some of it falling back towards the AGN disc, while other parts may escape from the system. Therefore, several observational effects can be anticipated: (i) the initial explosion is similar to that of a normal SN Ia, being powered by radioactivity decay and peaking in optical band \citep{2013A&A...554A..27P}; (ii) the collision between the fallback ejecta and the AGN disc can give rise to the emergence of hard X-ray and soft gamma-ray transients. 
(iii) the destruction of the AGN disc by the fallback ejecta results in a temporary increase, then decrease, and finally a return to normal levels again in the accretion rate and luminosity \citep{2022MNRAS.514.4102M}, similar to the observed changing-look AGN \citep{ricci_destruction_2020, 2019ApJ...883...94T, 2022ApJ...933...70L}. 

The collisions between the fallback ejecta and the AGN disc, in conjunction with the supernova explosions of the WD system, give rise to significant new features. In the subsequent section, a model is constructed to explore the observational consequences resulting from these interactions (ii). The effects of (i) and (iii) will be discussed in detail in the Discussion section. 

\begin{table}
	\centering
	\caption{The mass of disc gas swept by ejecta as it rushes out of the disc. It is assumed that the explosions occur in the midplane of the AGN disc. The AGN disc model is taken from the standard disc model of \citet{kato1998black}.}
	\label{tab:table1}
	\begin{tabular}{lcccc} 
		\toprule[1pt]
		\toprule[1pt]
		\multicolumn{4}{c}{$M_{\rm SMBH}=10^6M_\odot$, $\dot{M}=0.1\dot{M}_{\rm Edd}, \alpha=0.1$} \\
		\midrule[0.5pt]
		Explosion radius ($r_{\rm g}$) & 100 & 200 & 500 \\
		\midrule[0.5pt]
		Swept disc mass $(\rm g)$ & $2.51\times10^{27}$ & $7.78\times10^{27}$ & $3.31\times10^{28}$ \\
		\toprule[1pt]
		\toprule[1pt]
		\multicolumn{4}{c}{$M_{\rm SMBH}=10^7M_\odot$, $\dot{M}=0.1\dot{M}_{\rm Edd}, \alpha=0.1$} \\
		\midrule[0.5pt]
		Explosion radius ($r_{\rm g}$) & 100 & 200 & 500 \\
		\midrule[0.5pt]
		Swept disc mass $(\rm g)$ & $2.51\times10^{29}$ & $7.78\times10^{29}$ & $3.31\times10^{30}$ \\
		\toprule[1pt]
		\toprule[1pt]
		\multicolumn{4}{c}{$M_{\rm SMBH}=10^8M_\odot$, $\dot{M}=0.1\dot{M}_{\rm Edd}, \alpha=0.1$} \\
		\midrule[0.5pt]
		Explosion radius ($r_{\rm g}$) & 100 & 200 & 500 \\
		\midrule[0.5pt]
		Swept disc mass $(\rm g)$ & $2.51\times10^{31}$ & $7.78\times10^{31}$ & $3.31\times10^{32}$ \\
		\bottomrule[1pt]
	\end{tabular}
\end{table}
\subsection{Fallback model}
\label{sec:2.3}
As shown in Fig. ~\ref{fig:katong12}, it is assumed that the supernova explosion occurs in the midplane of the disc, resulting in symmetric motion of the ejecta about the disc plane. When two WDs collide, the center of mass moves in a circular Keplerian orbit around the SMBH. Following the triggered nuclear detonation after the collision, the ejecta undergoes homologous expansion with a typical velocity of $1.5\times 10^4 \rm\ km\,s^{-1}$. Subsequent to the supernova explosion, the ejecta is propelled out of the AGN disc at nearly its initial velocity. Under the influence of the SMBH gravity, certain portions of the ejecta fall back along the negative X-axis of the disc, while other portions escape from the system. The focus of interest lies in the radiation generated by the interaction between the fallback ejecta and the disc. A typical explosion radius of $\sim100r_{\rm g}$ is assumed, and Newtonian gravity is employed. It is worth noting that if the ejecta has the potential to fall back, the fallback points are situated on the X-axis, as depicted in the right panel of Fig. ~\ref{fig:katong12}. Furthermore, even some components of gravitationally unbound ejecta can fall back to the disc if their hyperbolic trajectory intersects the disc.

\begin{table}
	\centering
	\caption{Ratio of fallback ejecta with different model parameters. The columns from left to right in the table represent the mass of the SMBH, explosion radius, explosion velocity, the corresponding ratio of fallback ejecta.}	
	\label{tab:table2}
	\begin{tabular}{lcccc} 
		\toprule[1pt]
		\toprule[1pt]
		$M_{\rm SMBH}$ & $R_{\rm ej}$ & $v_{\rm ej}$ & Ratio \\
		\midrule[1pt]
		$10^6M_\odot$ & 100$r_{\rm g}$ & $1.5\times 10^4\rm\ km\,s^{-1}$ & 0.8873 \\
		\midrule[0.5pt]
		$10^6M_\odot$ & 200$r_{\rm g}$ & $1.5\times 10^4\rm\ km\,s^{-1}$ & 0.7107 \\
		\midrule[0.5pt]	
		$10^6M_\odot$ & 300$r_{\rm g}$ & $1.5\times 10^4\rm\ km\,s^{-1}$ & 0.6204 \\
		\midrule[0.5pt]	
		$10^6M_\odot$ & 500$r_{\rm g}$ & $1.5\times 10^4\rm\ km\,s^{-1}$ & 0.5121 \\
		\midrule[0.5pt]
		$10^6M_\odot$ & 100$r_{\rm g}$ & $1.0\times 10^4\rm\ km\,s^{-1}$ & 1 \\
		\midrule[0.5pt]	
		$10^6M_\odot$ & 100$r_{\rm g}$ & $2.0\times 10^4\rm\ km\,s^{-1}$ & 0.7385 \\
		\midrule[0.5pt]
		$10^7M_\odot$ & 100$r_{\rm g}$ & $1.5\times 10^4\rm\ km\,s^{-1}$ & 0.8873 \\
		\midrule[0.5pt]	
		$10^8M_\odot$ & 100$r_{\rm g}$ & $1.5\times 10^4\rm\ km\,s^{-1}$ & 0.8873 \\
		\bottomrule[1pt]
	\end{tabular}
\end{table}
Specifically, in our model the ejecta is assumed to be distributed spherically symmetrically at the explosion point. The ejecta is divided into $N$ parts, with $N=2\times10^5$. Monte Carlo simulations are conducted to determine the polar angle and azimuth angle of the explosion ejecta. For each part of the ejecta, the initial position and velocity are taken as the position and velocity of the ejecta at the explosion point, respectively. The trajectory of each part of the ejecta can then be determined, allowing us to determine whether it falls back to the disc, and, if so, obtain its time, position, and velocity. Since the fallback ejecta is typically dispersed widely upon reaching the disc, the mass of the fallback ejecta per $r_{\rm g}$ is lower than the mass of the disc gas per $r_{\rm g}$ (with exceptions discussed in Section 3). The impact of the fallback ejecta on the fast-rotating disc gas leads to the generation of shock waves, which efficiently convert kinetic energy into internal energy, thereby resulting in the production of electromagnetic radiation. It is assumed that, once the fallback ejecta has swept a mass equal to itself in the disc, all the kinetic energy is converted into internal energy and the dissipated energy for each part of the ejecta can be expressed as
\begin{equation}
	E_{\rm i}=\dfrac{1}{2}\dfrac{M_{\rm ej}}{N}v_{\rm rel}^2,
	\label{eq:3}
\end{equation}
where $M_{\rm ej}$ represents the mass of the explosion ejecta and $v_{\rm rel}$ is the relative velocity between the fallback ejecta and the disc gas. It should be noted that different parts of the ejecta generally exhibit different $v_{\rm rel}$. However, only a portion of the energy can be released as electromagnetic radiation, and the efficiency is conservatively set to approximately $\eta\sim5\%$. This value encompasses the uncertainty of the model and ensures the validity of the qualitative analysis. Thus, for each part of the ejecta, only the energy $\eta\,E_{\rm i}$ is converted into electromagnetic radiation. This efficiency is similar to that assumed in the simulations conducted by \citet{2016ApJ...830..125J}, which studied the collisions of debris from a tidal disruption event (TDE). Although the kinetic energy of the debris stream is not completely converted into internal energy in these simulations, the final radiation efficiency is still 2$\%$\,-\,8$\%$. \citet{2004ApJ...610..368W} qualitatively estimated that the radiation energy resulting from a SN Ia explosion impacting an AGN disc is approximately $10^{50}\rm\ erg$. This radiation energy is equivalent to a radiation efficiency of $\sim$5$\%$.

\begin{figure*}
	\includegraphics[width=2.1\columnwidth]{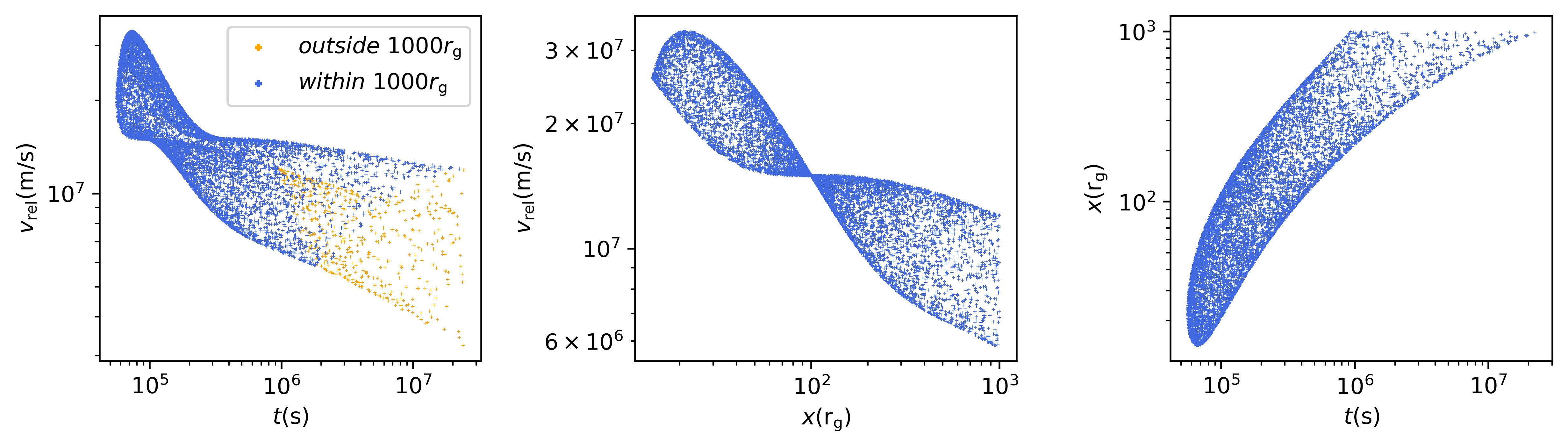}
	\caption{Physical properties of the ejecta that fall back onto the AGN disc in the fiducial model. The \textbf{left panel} gives the relationship between $v_{\rm rel}$ and the fallback time, the blue points represent the ejecta that fall back within $10^3r_{\rm g}$ while the orange points represent the ejecta that fall back outside $10^3r_{\rm g}$. The \textbf{middle panel} shows the relationship between the $v_{\rm rel}$ and the fallback position, and the \textbf{right panel} illustrates the relationship between the fallback position and the fallback time.}
	\label{fig:scatter}
\end{figure*}

Setting $M_{\rm SMBH}=10^7M_\odot$, explosion radius $R_{\rm ej}=100r_{\rm g}$, explosion velocity $v_{\rm ej}\sim1.5\times 10^4 \rm\ km\,s^{-1}$ and ejecta mass $M_{\rm ej}=1.8M_\odot$ as our fiducial model parameters, the corresponding fallback ejecta ratio (ratio of the total mass of fallback ejecta versus the total mass of ejecta) is 0.8873. The effects of varying the model parameters on the fallback ejecta ratios are shown in Table ~\ref{tab:table2}. Because the escape velocity decreases with increasing explosion radius, a larger explosion radius or explosion velocity yields a smaller fallback ratio. The SMBH mass does not affect the fallback ratio. For all cases considered here, more than half of the ejecta fall back towards the disc.

\section{Light curve}
\label{sec:3}
With a given set of model parameters for a supernova explosion, the fallback ratio, along with the ejecta falling information including the time, position, and velocity, can be obtained. This allows us to calculate the rate of ejecta fallback and the rate of energy dissipation with time, and hence the light curve.

The scatter diagram in Fig. ~\ref{fig:scatter} illustrates the 
physical properties of the fallback ejecta. The left panel shows the relationship between $v_{\rm rel}$ and the fallback time; the middle panel gives the relationship between $v_{\rm rel}$ and the fallback position; and the right panel shows the relationship between the fallback position and the fallback time. Each point represents one part of the fallback ejecta. In order to improve distinguishability, the scatter plot has been refined by using the results with $N=2\times10^4$. In the left panel, the orange points represent the ejecta falling back within $10^3\ r_{\rm g}$, while the blue points represent those that fall back outside $10^3\ r_{\rm g}$. During the early period, the relative velocity is widely dispersed, and can be up to $3.5\times10^7\rm\ m\,s^{-1}$. Therefore, the average dissipated energy for each ejecta is relatively large at this time. A the later time ($>10^6\rm\ s$), the relative velocity is distributed below $1.5\times10^7\rm\ m\,s^{-1}$. As time goes by, the distribution of the relative velocity changes slowly, and the average dissipated energy for the ejecta decreases gradually. Similarly, in the middle panel, the ejecta falling at $\lesssim100r_{\rm g}$, the relative velocity is more dispersed, the average relative velocity is larger, and so each part of the ejecta dissipates more energy. At $\gtrsim100r_{\rm g}$, the average relative velocity becomes smaller and the energy dissipation for each part of the ejecta is less. Generally speaking, for the farther fallback
position the relative velocity is found to be lower. At $\sim100r_{\rm g}$, there is a point where the relative velocity is concentrated at $\sim1.5\times10^7\rm\ m\,s^{-1}$. From the right panel, it can be seen that, during the early period, fallback positions are in the inner region of the disc ($\sim15r_{\rm g}$). Later on, the fallback positions disperse rapidly, and the average distance of the fallback positions increases rapidly as well. This is because a longer time of movement is usually required for points falling farther from the SMBH. In general, the fallback ejecta are dispersed in time and space.

In the following, we first examine the effects of the SMBH mass, while keeping the other parameters fixed to those of the fiducial model. At $100r_{\rm g}$, the Keplerian velocity is $c/10$, which is greater than the explosion velocity of a supernova. Therefore, in the fiducial model, all ejecta is still along the side of the moving direction of the center of mass (i.e. the Y-component of all ejecta velocities is positive). It is also worth noting that the relative velocities for collisions with $c/10$ have a relativistic correction of less than $1\%$.
This indicates that the relativistic effects of ejecta are relatively negligible in this scenario.

\begin{figure}
	\includegraphics[width=\columnwidth]{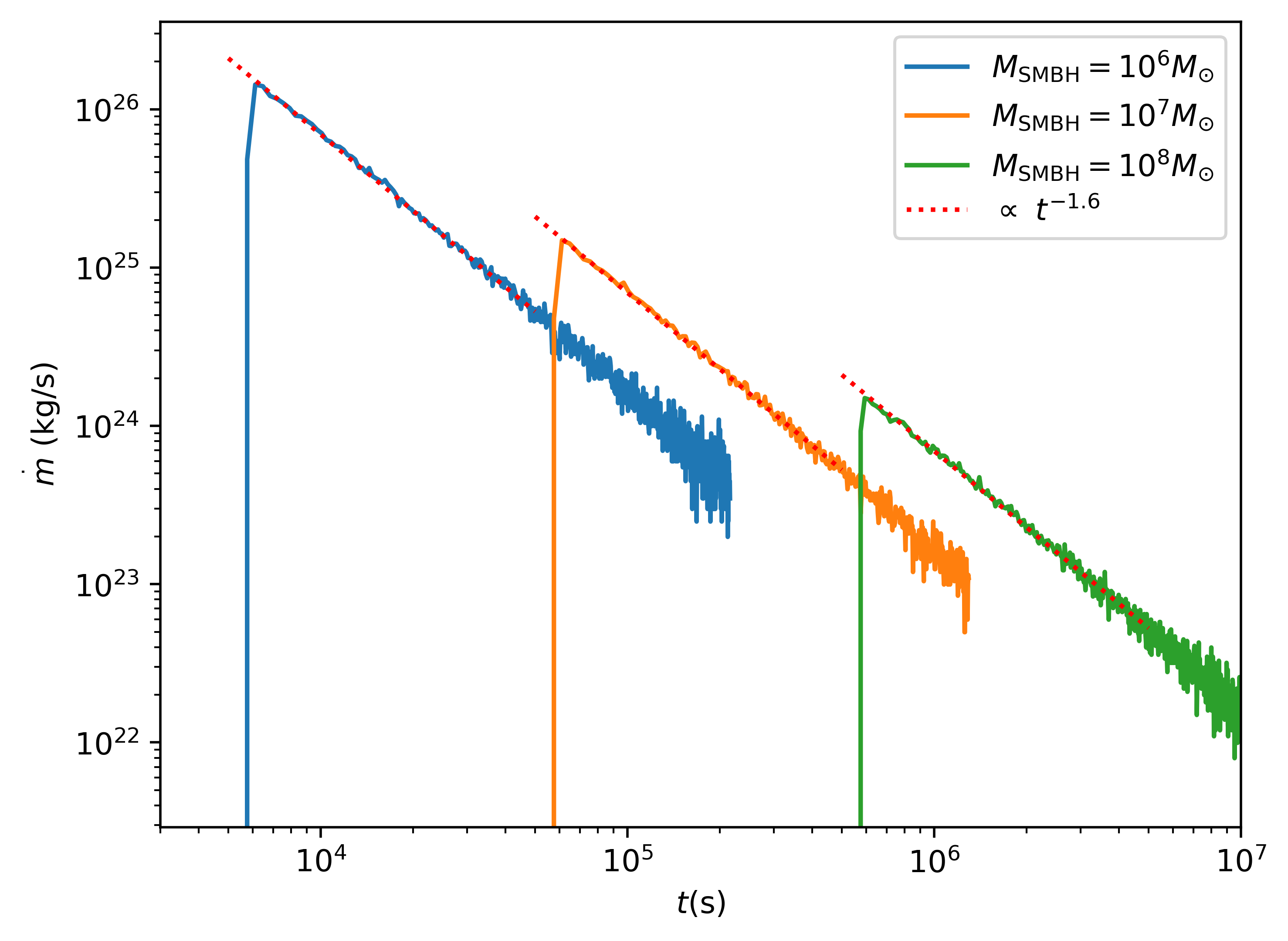}
	\caption{The rate of the fallback mass as a function of time. Different solid lines correspond to the cases with the different SMBH masses. The red dotted line follows $L\propto\,t^{-1.6}$. The rest of the parameters are taken from the fiducial model.}
	\label{fig:massrateVSt3}
\end{figure}

Fig. ~\ref{fig:massrateVSt3} shows the rate of the fallback rate as a function of time. Different lines correspond to the different SMBH masses, which are $M_{\rm SMBH}=10^6M_\odot$,$10^7M_\odot$ and $10^8M_\odot$, respectively. The profiles of different lines are almost the same, but the fallback time-scale is different due to the different central SMBH masses. Since the total ejecta masses are the same, that is, the amount of the material ejected from a SN Ia explosion, the peak fallback rate decreases by an order of magnitude as the SMBH mass increases by an order of magnitude. The jaggedness in the curves results from the finite number used in the simulations ($N=2\times10^5$), which can be improved by increasing $N$. 

\begin{figure}
	\includegraphics[width=\columnwidth]{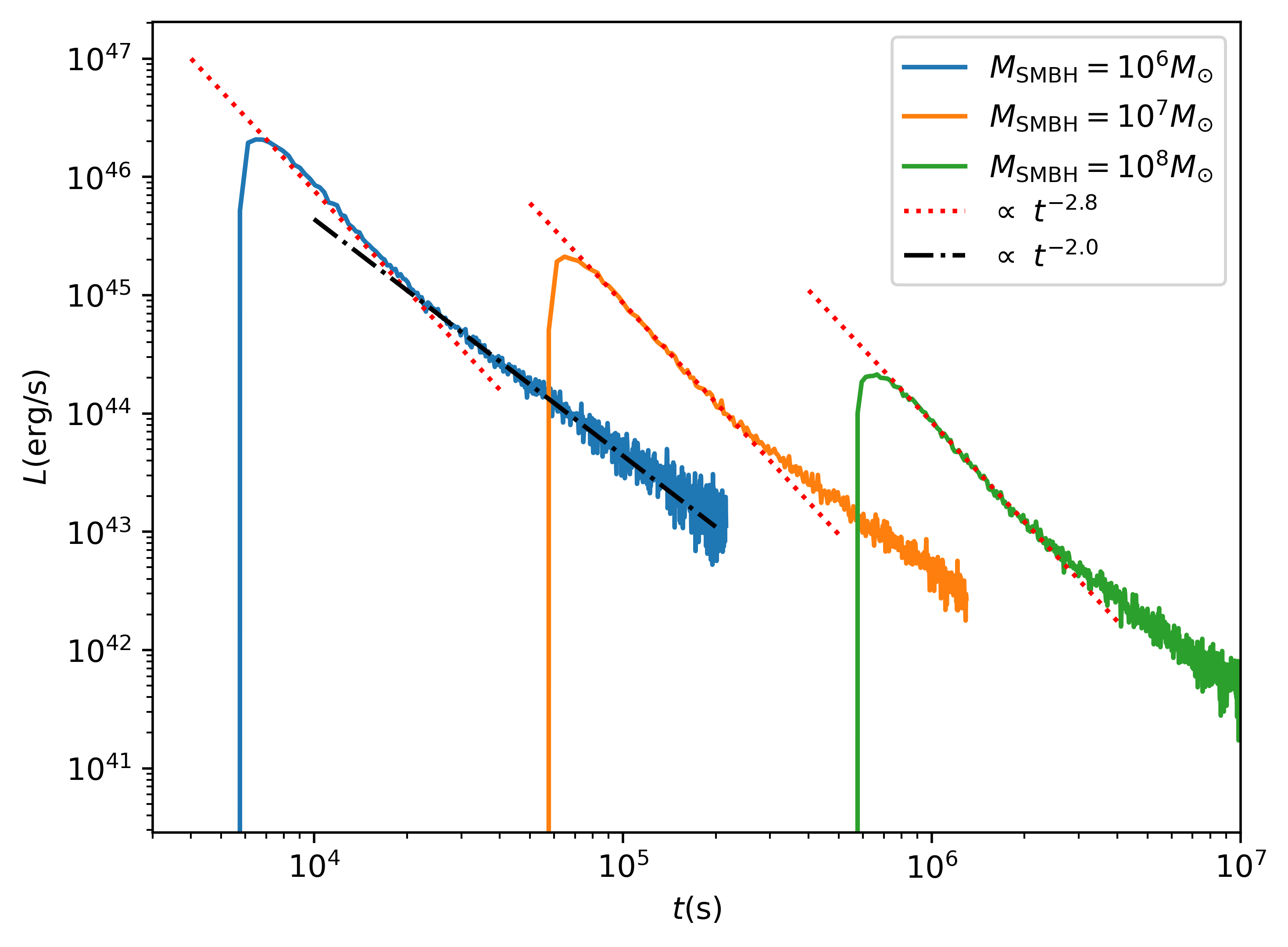}
	\caption{The light curves originating from the energy dissipation caused by fallback ejecta hitting the disc gas. The red dotted lines and black dashed line follow $L\propto\,t^{-2.8}$ and $L\propto\,t^{-2.0}$, respectively. The rest of the parameters are the same as these of the fiducial model.}
	\label{fig:LVSt3}
\end{figure}

Fig. ~\ref{fig:LVSt3} shows the light curves. The profiles of the light curves are similar to the time evolution of the rate of the matter fallback, characterized by a fast rise and then followed by a power-law decay. However, the light curve is related not only to the fallback rate but also to the corresponding $v_{\rm rel}$ and parameter $\eta$ in our model. For comparison, the curves $L\propto\,t^{-2.8}$and $L\propto\,t^{-2.0}$ are also plotted. In the initial period after the peak in the light curve, the luminosity decreases rapidly with a power law $L\propto\,t^{-2.8}$, and then with a moderate power law $L\propto\,t^{-2.0}$ in the subsequent period. The overall decline index is high, meaning the luminosity changes rapidly. Different SMBH masses do not affect the light-curve profile, but they do influence the time-scale of variability and peak luminosity. As SMBH mass increases by an order of magnitude, the time-scale of variability also increases an order of magnitude, and the peak luminosity decreases by an order of magnitude. These trends offer the possibility of using the variability timescale to constrain the mass of the SMBH.

\begin{figure}
	\includegraphics[width=\columnwidth]{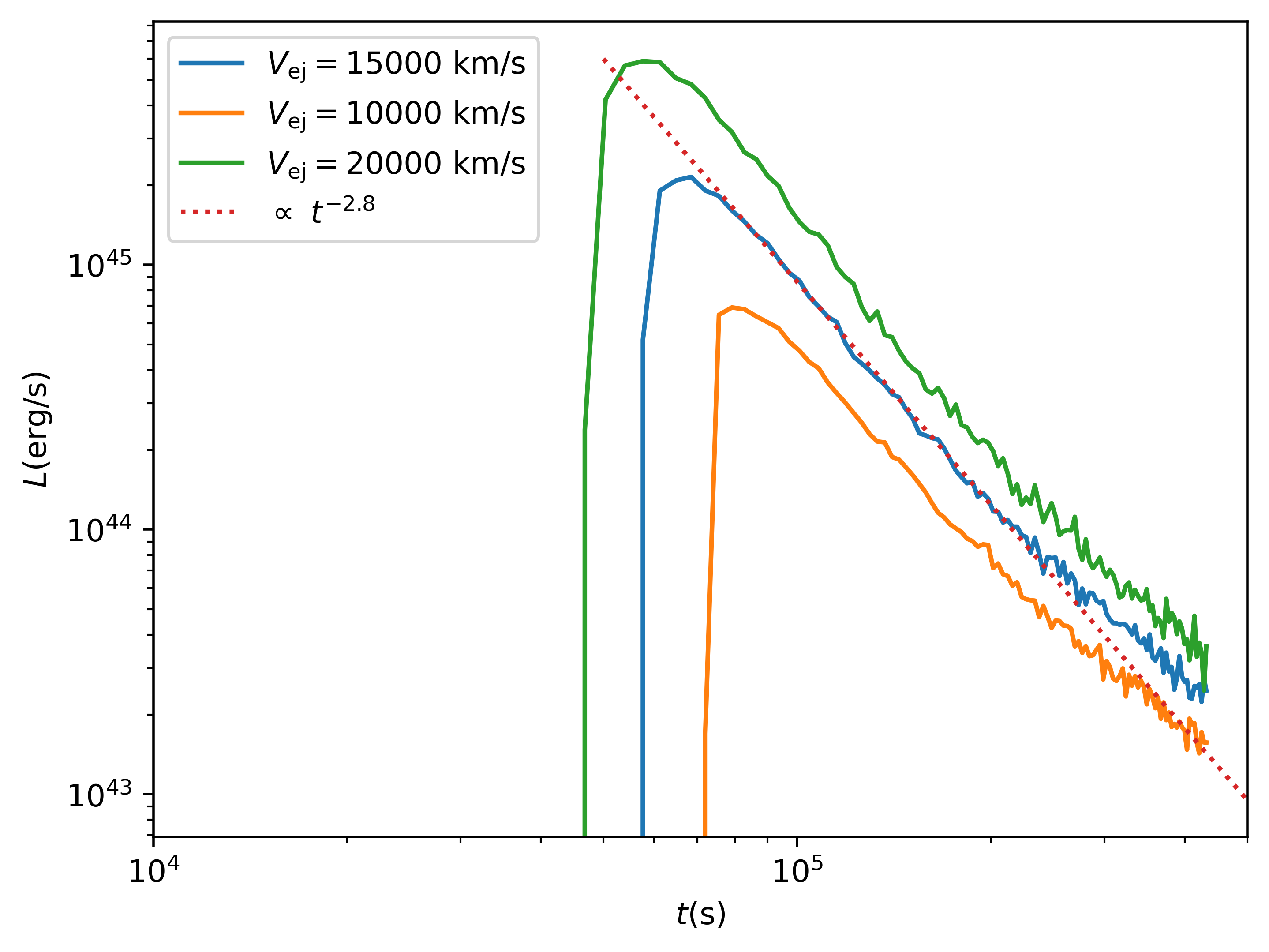}
	\caption{Light curves for different explosion velocities, compared with the fiducial model and $L\propto\,t^{-2.8}$.}
	\label{fig:Vej123}
\end{figure}

\begin{figure}
	\includegraphics[width=\columnwidth]{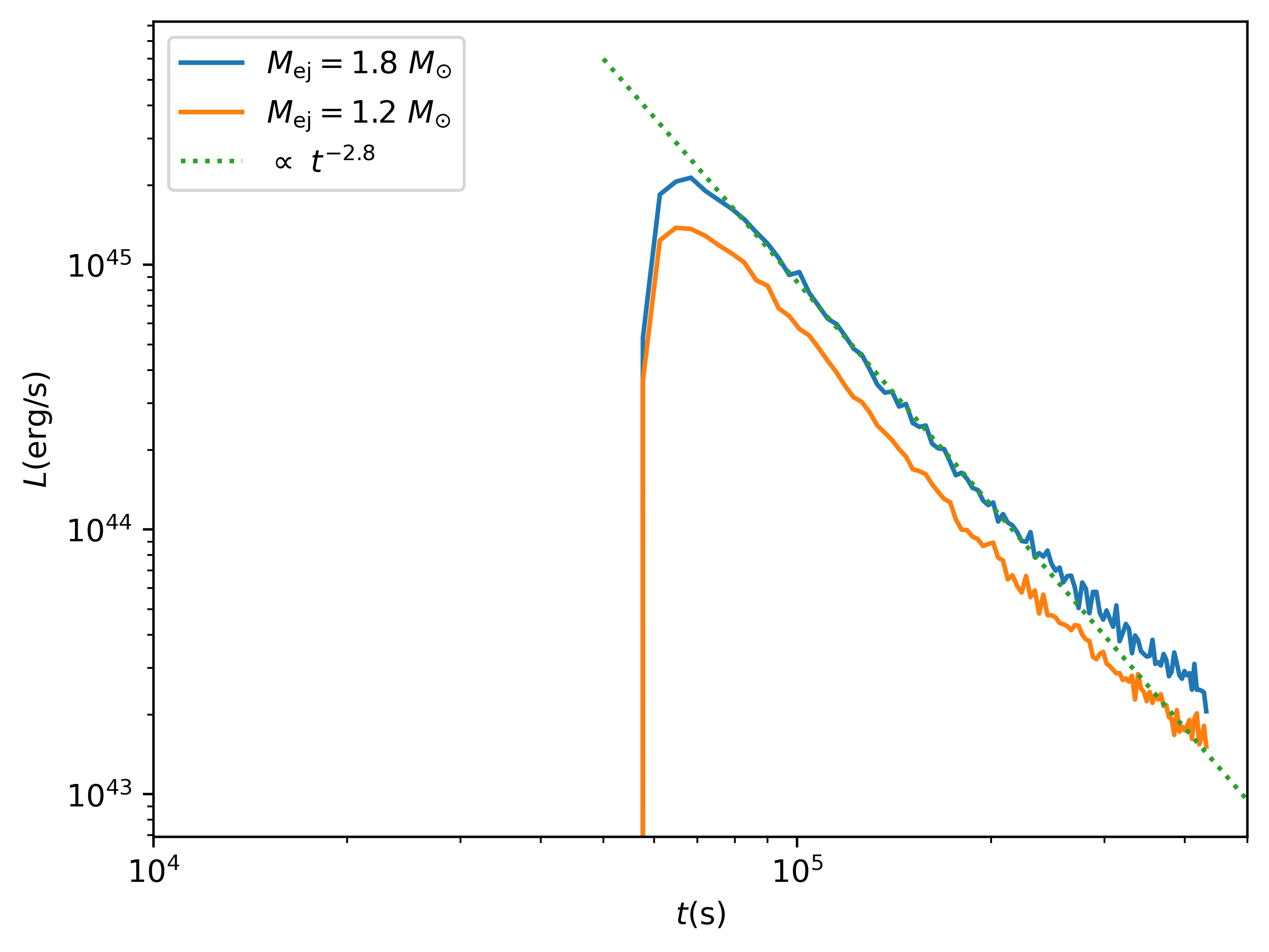}
	\caption{Light curve for different ejecta mass, compared with the fiducial model and $L\propto\,t^{-2.8}$.}
	\label{fig:Mej12}
\end{figure}

\begin{figure}
	\includegraphics[width=\columnwidth]{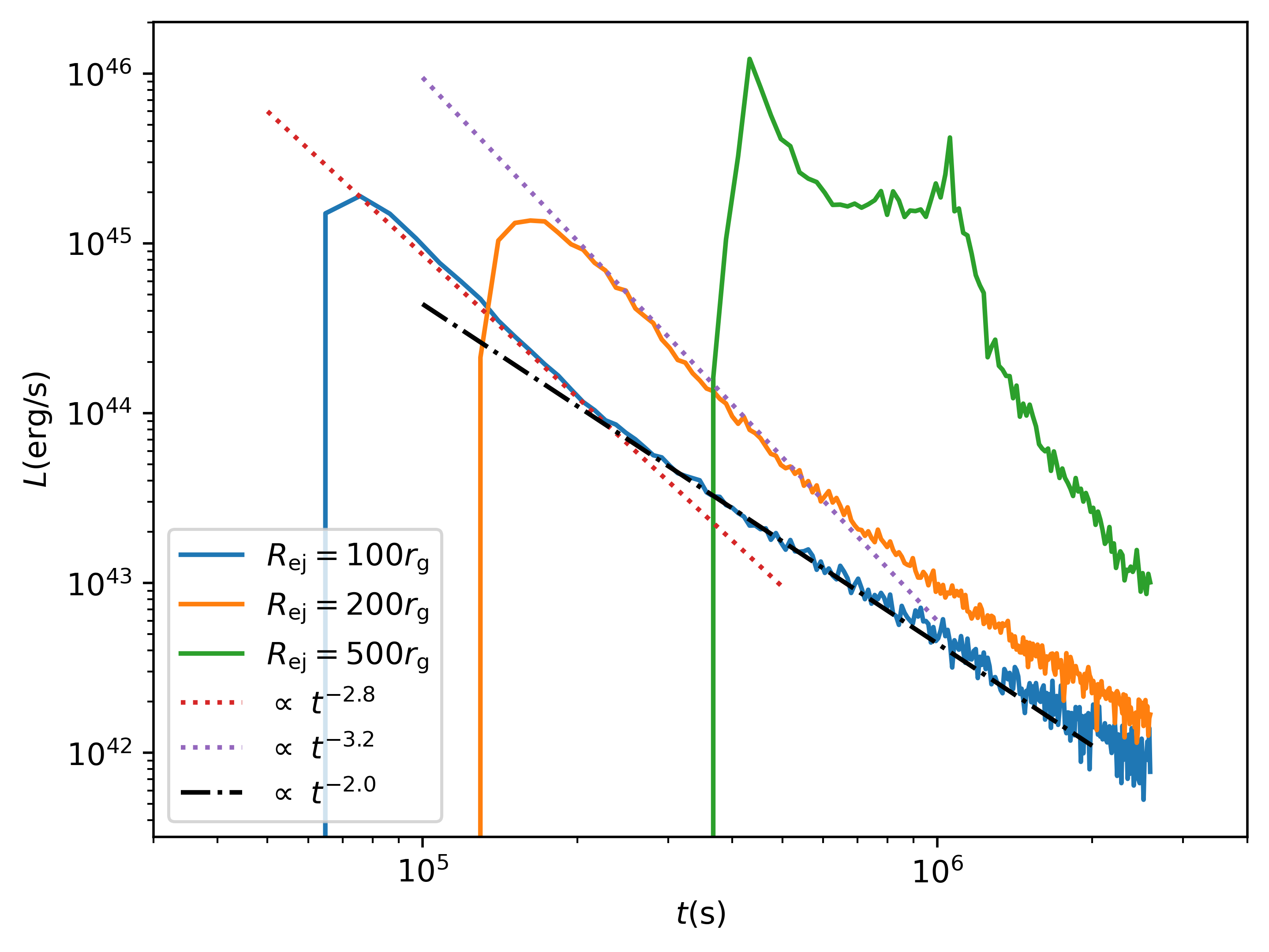}
	\caption{Light curves for different explosion positions, compared with the fiducial model and $L\propto\,t^{-2.8}$.}
	\label{fig:Rej123}
\end{figure}

The effects of different parameters from the fiducial model are shown further in Figs ~\ref{fig:Vej123} -- ~\ref{fig:Rej123}. The light curves of different explosion velocities are compared in Fig. ~\ref{fig:Vej123}. It can be seen from Table ~\ref{tab:table2} that the fallback ratio of ejecta decreases with increasing explosion velocity. From the light curves, however, it is found that a larger explosion velocity, resulting in higher $v_{\rm rel}$, can lead to an earlier peak time and a larger peak luminosity, although the power-law shape of the light curve remains unchanged. A lower explosion velocity ($v_{\rm rel}=10^4\rm\ km\,s^{-1}$) means that the ejecta is affected by the AGN disc and partially is decelerated during the initial phase; or it can mean that the explosion result deviates from the ideal collision with zero impact parameter \citep{2009MNRAS.399L.156R}. The light curves for the cases of the different ejecta masses are compared in Fig. ~\ref{fig:Mej12}. Different ejecta masses can only change the peak luminosity while keeping the peak time and the profile of light curves unchanged. 

The light curves for the cases of the  different explosion radii are compared in Fig. ~\ref{fig:Rej123}. Different explosion radii affect the profile of the light curve. It is worth noting that the green curve corresponding to the explosion radius at $500r_{\rm g}$ has two peaks. This is because the Kepler velocity at $500r_{\rm g}$ is lower than the explosion velocity. In contrast to the explosion at $100r_{\rm g}$, the ejecta after the explosion  still moves along the direction of the side of the center of mass, but some parts of the ejecta along the opposite side in $500r_{\rm g}$ case (i.e. the Y-component of some ejecta velocities is negative.) These parts of the ejecta produce two peaks in the light curve. For the larger explosion radii or larger explosion velocity, the fallback ratio is smaller (Table ~\ref{tab:table2}). However, it is obvious that the peak value of the light curve is not determined by the fallback ratio alone. Generally, two physical factors affect the power-law slopes of the light curve: the fallback rate of ejecta and the change rate of the average relative velocity between the fallback ejecta and the disc gas. In the case of a single light curve, while the fallback rate of ejecta varies with time following a power-law slope (as shown in Fig. ~\ref{fig:massrateVSt3}), the relative velocity does not vary with time following a power-law slope (as shown in Fig. ~\ref{fig:scatter}). Therefore, the change in the average relative velocity over time is the factor responsible for the different power-law slopes of the same light curve at different times. When comparing different light curves, their fallback rate of matter and the change rate of average relative velocity differ, resulting in varying curve profiles and power-law slopes.


\begin{figure}
	\includegraphics[width=\columnwidth]{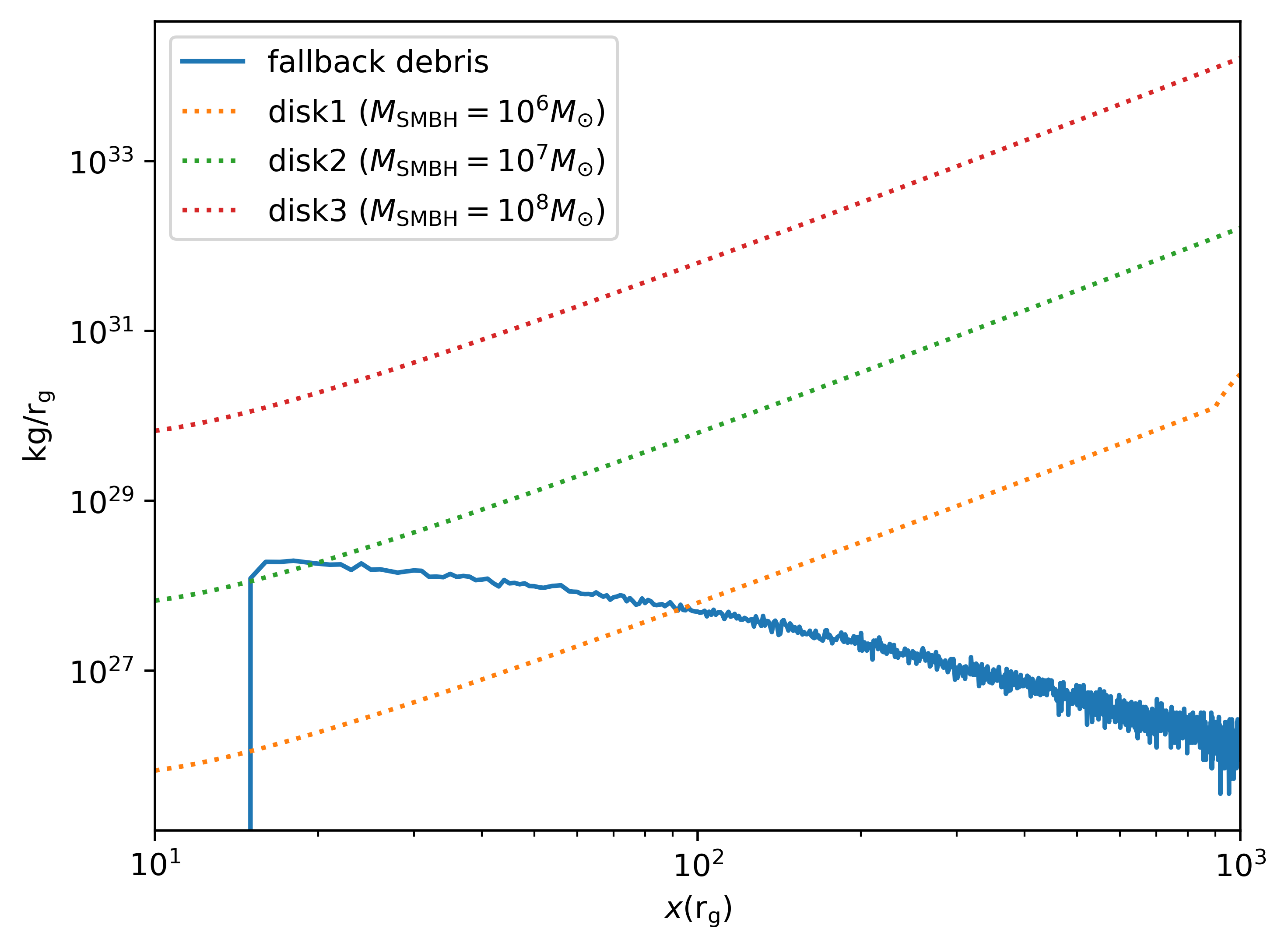}
	\caption{The time-integrated distribution of the fallback ejecta in AGN disc (blue curve) compared with the different distribution of AGN disc gas (orange, green and red dotted curves represent $M_{\rm SMBH}=10^6M_\odot$, $10^7M_\odot$, $10^8M_\odot$ respectively). The blue curve show the result of the case with the same fiducial parameters but the different masses of the central SMBH.}
	\label{fig:massVSx}
\end{figure}

Several assumptions are made in the above calculations. When the ejecta falls back onto the AGN disc, it interacts primarily with the surface material of the disc. Radiation can escape the disc promptly, since the optical depth in this location is not too large (see Section \ref{sec:4} for estimates). It is assumed that the fallback ejecta is completely absorbed by the disc gas and that the kinetic energy is converted into internal energy, with efficiency $\eta \sim 5\%$ converting internal energy into electromagnetic radiation. However, as shown in Fig. ~\ref{fig:massVSx}, this assumption does not hold in the inner part of the AGN disc: for $M_{\rm SMBH}=10^6M_\odot$, the falling ejecta stream is heavy enough to penetrate the disc.
The term 'heavy' means that the time-integrated fallback ejecta per unit radius is greater than the disc gas per unit radius; this should not be confused with 'heavy element'. In Fig. ~\ref{fig:massVSx}, the blue curve represents the time-integrated distribution of the fallback ejecta in the AGN disc, which is compared with the different distributions of AGN disc gas. The blue curve is taken from the fiducial parameters, but it is suitable for the different central SMBH masses. The heavy falling stream can affect the stable structure of the AGN disc. We do not quantitatively consider the destruction of the disc and subsequent effects further, but such processes should be correlated with changing-look AGN, as described in the model for the source 1ES 1927+654 \citet{2019ApJ...883...94T, ricci_destruction_2020, 2022ApJ...933...70L}. Therefore, the peak range of the actual light curve should deviate from the blue curve of Fig. ~\ref{fig:massrateVSt3}. However, it is still reliable and can approximate the light curve at this time. Because the fallback ejecta is symmetrical about the AGN disc plane, the velocity in the $Z$-direction can cancel out, and only a small part of the kinetic energy is consumed by the AGN disc.

\section{Spectral energy distribution}
\label{sec:4}
Suppose that after collisions between the fallback ejecta and AGN disc gas with equal masses, all kinetic energy is converted into thermal energy through shocks. Radiation of the thermal energy can be approximated with a blackbody spectrum. The optical depth at the shock location can be estimated as, 
\begin{equation}
	\tau\sim\sigma\dfrac{\Sigma}{2m_{\rm H}}, 
	\label{eq:ot}
\end{equation}
here, $\sigma$ represents the cross section of photons, and $\Sigma$ denotes the average column density of the fallback ejecta, which can be estimated from Fig. ~\ref{fig:massVSx}. The cross-section for radiation can be written as \citep{rybicki1991radiative},
\begin{equation}
\begin{split}
	\sigma=&\dfrac{3}{4}\sigma_{\rm T}\left\{\dfrac{1+\Gamma}{\Gamma^3}\left[\dfrac{2\Gamma(1+\Gamma)}{1+2\Gamma}-\ln(1+2\Gamma)\right]+ \right. \\ & \left.\dfrac{\ln(1+2\Gamma)}{2\Gamma}-\dfrac{1+3\Gamma}{(1+2\Gamma)^2}\right\},
	\label{eq:rati}
\end{split}
\end{equation}
where $\sigma_{\rm T}$ is the Thomson cross section and $\Gamma$ is the ratio of photon energy to the rest energy of electron, which is defined as
\begin{equation}
	\Gamma=\dfrac{h\nu}{m_{\rm e}c^2},
	\label{eq:Gamma}
\end{equation}
where $h$ is the Planck constant, $m_{\rm e}$ is the rest mass of electron, and $\nu$ is the photon frequency.
In our fiducial model, the fallback location is at $\sim100r_{\rm g}$ ($\Gamma\sim10$), the thickness is $\tau \sim80\gtrsim1$, and thus the blackbody spectrum approximation is reasonable. The total SED during a period of time is determined by the superposition of all blackbody spectra.

Assuming that the average molecular weight of ejecta is $m_{\rm A}=Am_{\rm H}$, and considering that the AGN disc comprises ionized hydrogen, the temperature $T$ can be calculated using the equipartition theorem, 
\begin{equation}
	\dfrac{1}{2}m_{\rm A}v_{\rm rel}^2=\dfrac{3}{2}(1+A+\dfrac{3}{2}A)k_{\rm B}T.
	\label{eq:4}
\end{equation}
It follows that
\begin{equation}
	T=\dfrac{2Am_{\rm H}v_{\rm rel}^2}{3(2+5A)k_{\rm B}},
	\label{eq:5}
\end{equation}
where $k_{\rm B}$ is Boltzmann constant. When $A\gg1$,
\begin{equation}
	T\simeq\dfrac{2m_{\rm H}v_{\rm rel}^2}{15k_{\rm B}}.
	\label{eq:5r}
\end{equation}

\noindent For $M_{\rm SMBH}=10^6M_\odot$, if the fallback ejecta is located in the inner region of the AGN disc ($<100r_{\rm g}$), it is assumed that the fallback ejecta mainly collides with the symmetrical part of the fallback ejecta, resulting in
\begin{equation}
	\dfrac{1}{2}m_{\rm A}v_{\rm rel}^2=\dfrac{3}{2}(2+A)k_{\rm B}T,
	\label{eq:6}
\end{equation} 
and the temperature is
\begin{equation}
	T=\dfrac{Am_{\rm H}v_{\rm rel}^2}{3(2+A)k_{\rm B}}. 
	\label{eq:7}
\end{equation} 
When $A\gg1$,
\begin{equation}
	T\simeq\dfrac{m_{\rm H}v_{\rm rel}^2}{3k_{\rm B}}.
	\label{eq:7r}
\end{equation}

\begin{figure}
	\includegraphics[width=\columnwidth]{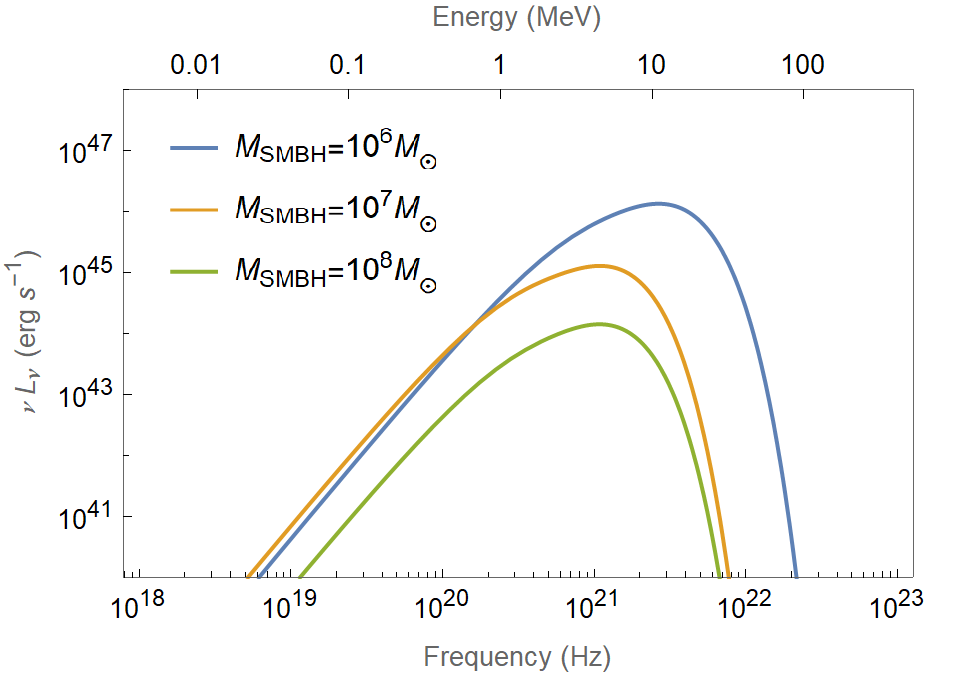}
	\caption{The SEDs at the peak luminosity for different SMBH masses. The rest of the parameters are taken from the fiducial model. The multi-temperature black body spectra are created by overlaying the SEDs generated by each part of the fallback ejecta.}
	\label{fig:SEDM678}
\end{figure}

Fig. ~\ref{fig:SEDM678} illustrates the SED at the peak luminosity for the cases of the different SMBH masses. It can be observed that the peak luminosity increases as the SMBH mass decreases, which is consistent with the results presented in Fig. ~\ref{fig:LVSt3}. All these SEDs exhibit multi-temperature blackbody spectra, with peak frequencies falling within the soft $\gamma$ range. Notably, for $M_{\rm SMBH}=10^6M_\odot$, the SED peaks at a relatively higher frequency. This is because in, the case of a less massive central SMBH, the disc gas in the inner region is not sufficiently dense and thus the fallback ejecta dissipates energy primarily through collisions with the symmetrical part of the ejecta. As a result, radiation of relatively higher energy is produced. For SMBHs of other masses ($M_{\rm SMBH}\gtrsim 10^7M_\odot$), the energy is dissipated by collisions between the fallback ejecta and the disc gas. This difference can be seen by comparing Equations (~\ref{eq:5}) and (~\ref{eq:7}). According to the equipartition theorem, the average energy of each particle obtained from kinetic energy dissipation is different, because the composition of the fallback ejecta is different from that of the AGN disc gas. However, the definition of $v_{\rm rel}$ in Equations (~\ref{eq:4}) and (~\ref{eq:6}) is the same (the relative velocity between the fallback ejecta and the disc gas). Since the fallback ejecta exhibits symmetry with respect to the AGN disc, the Z-component of the velocity cancels out and only a small portion of the remaining kinetic energy is transferred to the AGN disc.

\begin{figure}
	\includegraphics[width=\columnwidth]{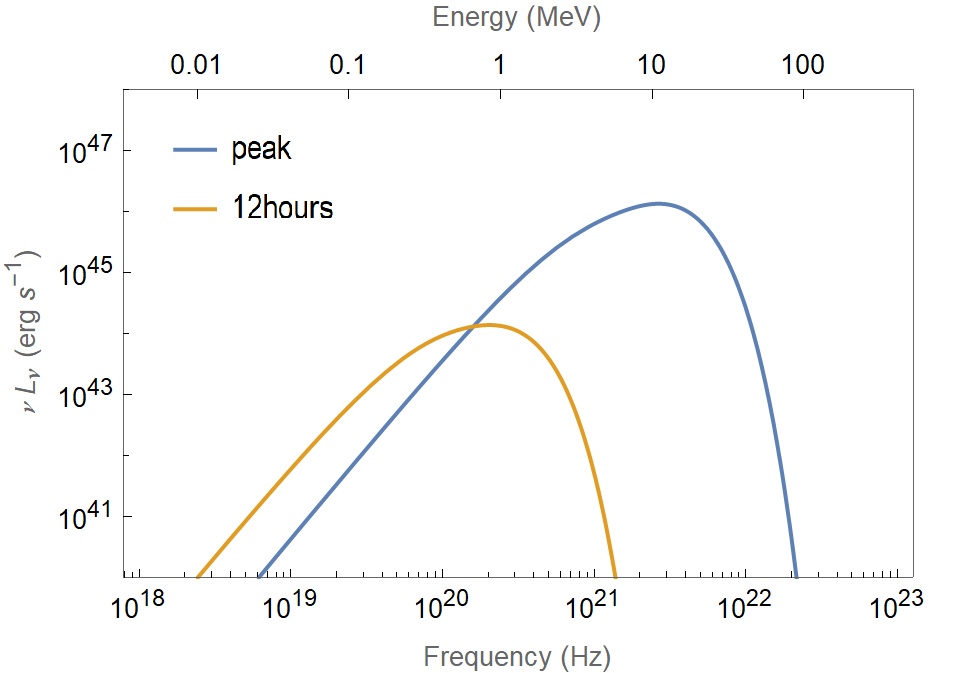}
	\caption{The SEDs at different times for $M_{\rm SMBH}=10^6M_\odot$. The blue curve corresponds to the phase when the light curve is at its peak, and the orange curve corresponds to $12$ hours after the explosion. The rest of the parameters are taken from the fiducial model.}
	\label{fig:SEDM6t12}
\end{figure}

\begin{figure}
	\includegraphics[width=\columnwidth]{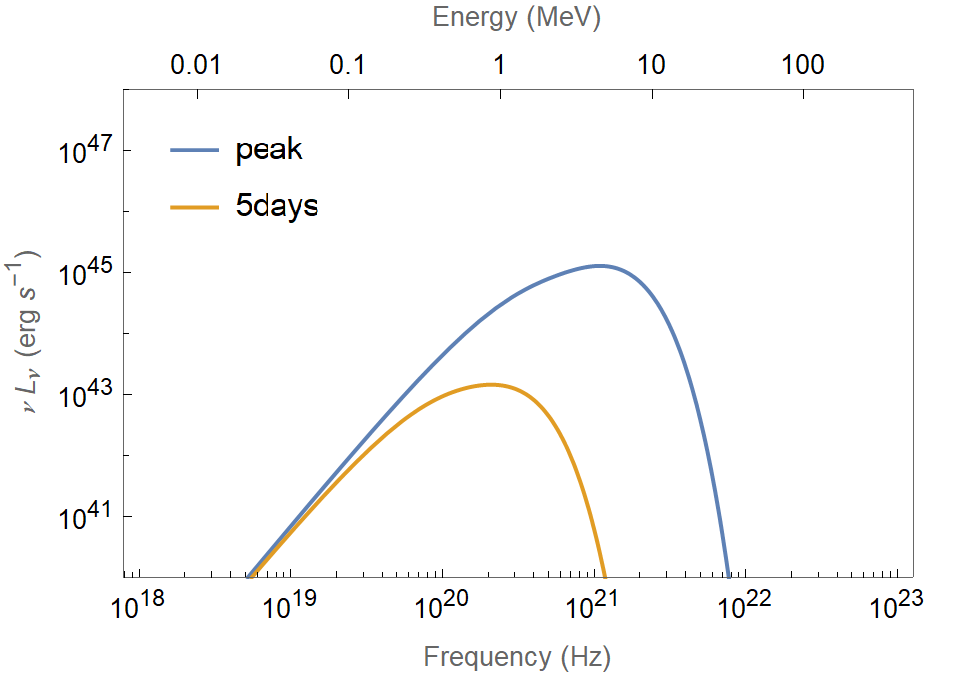}
	\caption{The SEDs at different times for the fiducial model. The blue curve corresponds to the phase when the light curve is at its peak, and the orange curve corresponds to $5$ days after the explosion.}
	\label{fig:SEDM7t12}
\end{figure}

Figs ~\ref{fig:SEDM6t12} and ~\ref{fig:SEDM7t12} show the SEDs 
for the cases of $M_{\rm SMBH}=10^6M_\odot$ and $M_{\rm SMBH}=10^7M_\odot$, respectively. The lines in the different colors represent the SED at the different times. At the maximum of the light curve, the peak frequency of the SED is in the soft $\gamma$ range due to the large $v_{\rm rel}$, but at late times the peak frequency shifts into the hard X-ray range. Except for some blazars, the average luminosities of AGNs at $\rm MeV$ energies are $\lesssim10^{40}\rm\ erg\,s^{-1}$ \citep{2001ApJ...559..680H}. Therefore, the high-energy radiation caused by the fallback ejecta of SNe Ia when it hits the disc gas far exceeds the AGN background luminosity and it should be detectable.

\begin{figure}
	\includegraphics[width=\columnwidth]{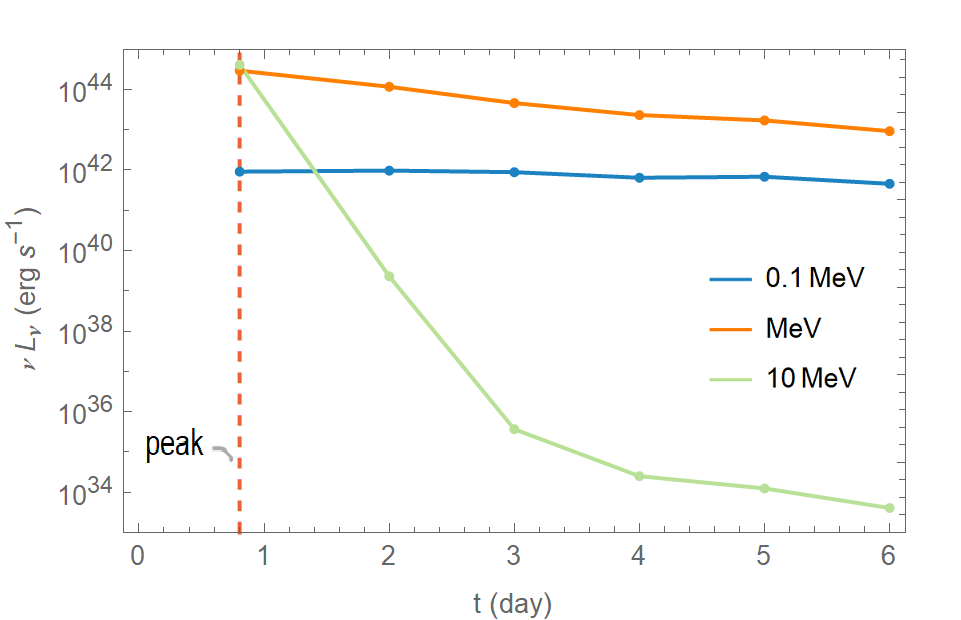}
	\caption{The light curves in different energies around the $\rm MeV$ range for the fiducial model. The red dotted line indicates the peak time of the total luminosity, and the light curves in different energies are showed from the peak time.}
	\label{fig:vLvcurve}
\end{figure}

Fig. ~\ref{fig:vLvcurve} shows the light curves in different energies around the $\rm MeV$ range for the fiducial model. The red dotted line indicates the peak time of the total luminosity, and the light curves in different energies are shown from the peak time. It can be seen from the figure that the higher the energy, the faster the curve varies. Although the peak luminosity of $10\rm MeV$ is very high, reaching about $\gtrsim10^{40}\rm\ erg\,s^{-1}$, it rapidly decreases by about 10 orders of magnitude within 6 days from the peak time; the luminosity of $1\rm MeV$ only decreases by about one order of magnitude within 6 days; while the luminosity of $0.1\rm MeV$ keeps almost as a constant within 6 days. Therefore, the high energy radiation of $\gtrsim 10\rm MeV$ is mainly concentrated near the peak time of the total luminosity, while the low-energy radiation can last for a relatively long time. These trends are the same for the SMBHs with the different masses, but the time-scales vary with the mass of the SMBH, correspondingly. We also notice that the luminosity 
in the $0.1\rm MeV$ band increases (rather than decreases) after the peak of the total luminosity for $M_{\rm SMBH}=10^6M_\odot$. This feature can also be seen in Fig. ~\ref{fig:SEDM6t12}, and is caused by the collisions between symmetrical fallback ejecta in the inner part of the AGN disc. 
In contrast, the light variation resulting from the initial explosion and radioactive decay of SNe Ia is not influenced by the mass of the SMBH. This is due to the fact that the half-life of radioactive elements involved in the decay process remains constant. 

For comparison, \citet{2004ApJ...610..368W} estimated a peak photon energy of approximately 400$\rm keV$ for SNe Ia ejecta impacting an AGN disc. Additionally, \citet{2021ApJ...914..107C} calculated that the fallback debris from a TDE colliding with an AGN disc generates a photon energy peak of up to $\sim\rm MeV$. A more comprehensive discussion on the limitations of the estimated SEDs has been provided in Section \ref{sec:5.1}.

The sky at MeV energies is currently poorly explored. From Fig. ~\ref{fig:vLvcurve}, it is evident that the luminosity in the fiducial model at an energy of $1\rm MeV$ is approximately $10^{43}\rm\ erg\,s^{-1}$. According to our estimation, \textit{Fermi} Gamma-ray Burst Monitor \citep[GBM:][]{2009ApJ...702..791M} could detect the gamma-ray emission at 1MeV resulting from the collision of binary WDs in AGNs within only 1 Mpc. However, some of our predicted events, with peak energies estimated to be around 10-50 MeV, might be detectable in \textit{Fermi} data using the Large Area Telescope (LAT) Low-Energy technique, which can significantly increase the effective area down to 30 MeV
\citep{2010arXiv1002.2617P}. Our predicted event may also be observed by future MeV detectors such as Compton Spectrometer and Imager (COSI) \citep{2022icrc.confE.652T}, Cube Sat MeV telescope (MeVCube) \citep{2022JCAP...08..013L}, e-ASTROGAM \citep{2018JHEAp..19....1D}, ComPare \citep{2022SPIE12181E..2GS,2015arXiv150807349M}, the All-sky Medium Energy Gamma-Ray Observatory eXplorer (AMEGO-X) \citep{Fleischhack:2021hJ}, the Galactic Explorer with a Coded aperture mask Compton telescope (GECCO) \citep{2022JCAP...07..036O}, and the Advanced Particle-astrophysics Telescope (APT) \citep{2022HEAD...1940404B}. Taking AMEGO-X as an example, its sensitivity at the $1\rm MeV$ energy is $5\times10^{-6} \rm MeV\,cm^{-2} s^{-1}$ \citep{Fleischhack:2021hJ}, so AMEGO-X can detect our predicted burst at $\sim 100 \rm Mpc$. This distance is much larger than the known nearest Seyfert galaxy \citep{2003ApJ...588L..13F}. 


\section{Discussion}
\label{sec:5}

\subsection{Model uncertainties}
\label{sec:5.1}

Because a significant WD population within the disc is a key facet of our model, it is necessary to discuss this assumption in greater detail. The initial mass of stars in the AGN disc environment can be 'top heavy' due to the high gas density in the disc, and the accretion of stars can change the distribution of the stellar mass further. However, it can be seen from fig. 7 of \citet{2022arXiv220510382D} that the peak value of stellar initial mass functions (IMFs) is under $8M_\odot$ except for $M_{\rm SMBH}=10^6M_\odot$, and a star below $8M_\odot$ can evolve into a WD. Even after $10^6$years of accretion, the peak value of the stellar mass function in an AGN disc with $M_{\rm SMBH}=10^6M_\odot$ can still reach below $8M_\odot$. Moreover, the stellar mass function is also related to the location in an AGN disc: for instance, the characteristic mass of protostar formed by in situ collapse is smaller in the inner region of the AGN disc (e.g. fig. 5 of \citet{2022arXiv220510382D}). 

It is noteworthy that the AGN disc lifetimes may be less than the main-sequence lifetimes of low-mass stars. 
According to the duty cycle model of AGNs \citep[e.g.][]{2009ApJ...690...20S}, 
although stars cannot evolve to WDs in one AGN activity cycle, they may evolve to WDs in the subsequent activity cycles or during periods of quiescence in the nucleus. 

Additionally, mass segregation in nuclear star clusters should cause WDs to occupy areas further from the centre compared with BHs or neutron stars (NSs), but not all nuclear clusters exhibit effective mass segregation \citep{mckernan_black_2020}. Around more massive SMBHs ($\gtrsim10^7M_\odot$), mass segregation may be less efficient and a less top-heavy IMF may be more appropriate. Such a circumstance should also apply to stars or compact objects that are captured into the disc.

For the capture of stars, their characteristic stellar mass can be lower than that of in situ collapse. While the accretion of captured stars in the inner region of an AGN disc increases the mass that could result in the formation of BHs, this is not the case in the outer region of the AGN disc \citep{2021ApJ...910...94C}. Following the accretion and evolution of stars, WDs formed in the outer region of the AGN disc can migrate to the inner region without undergoing collapse into BHs. In general, the IMF of stars in AGN discs tends to be 'top heavy,' with further mass increase through accretion. However, the initial mass, accretion, and evolution of stars are also influenced by their positions in the discs and the properties of the discs. Considering all these conditions, a significant number of WDs can still be generated in AGN discs.

Another facet of caution is the presence of migration traps. Although many works predict that migration traps exist in locations we are currently investigating, \citet{2020MNRAS.493.3732D} did not find any migration traps in their disc models. \citet{2021PhRvD.103j3018P} also suggested that including headwinds in migration leads to the complete disappearance of migration traps. Actually, our results do not necessarily rely entirely on migration traps, but we still expect the new observable phenomenon proposed in our paper to occur mainly at locations within $\lesssim10^3 r_{\rm g}$. The reasons are as follows. (i) Apart from migration traps, an overdensity of disc compact objects can also occur in regions of varying aspect ratios or varying migration time-scales \citep{2014MNRAS.441..900M}. According to \citet{2012MNRAS.425..460M}, the migration time-scale of $1-M_\odot$ compact stars increases inwards at $\lesssim10^3 r_{\rm g}$, although only the Type I migration torque is included. Alternatively, as depicted in fig. 3 of \citet{2021MNRAS.505.1324P}, the effects of GW torque and gas torque are considered. When the radius of the AGN disc decreases, the torque decreases or the migration timescale increases in the relative outer region of the AGN disc ($\sim10^3 r_{\rm g}$, which is not the trap region). Therefore, collisions may have occurred before WDs are migrated to traps due to the differential migration time-scale. (ii) Our simulations show that even if there are two WDs close to each other (the orbital separation of the two WDs is less than $2\sqrt{3}R_{\rm H}$), the probability of collision in the outer part of AGN disc can be reduced \citep{2023MNRAS.tmp.2113L}. (iii) Even if a small number of WDs collide in the outer region of AGN discs, the observational effect may not be obvious due to large amounts of the gas, unless the shock wave can break out on the disc surface \citep{2021ApJ...914L..19Z}. Even though the very external regions ($\gtrsim10^4 r_{\rm g}$) of a disc may initially be neutral, the early radiation from the explosion itself can ionize the neutral gas at a very short time-scale \citep{perna_electromagnetic_2021}. The ionization leads the disc to become optically thick, and the ejecta also can not rush out of the disc. In such cases, there might not be an overtly observable effect, but there is feedback on AGN discs. However, as noted by \citet{2022ApJ...928..191G}, the feedback is relatively insignificant compared to the accretion of compact objects. Alternatively, as in \citet{2021MNRAS.507..156G}, the ejecta can rush out of the disc if they are not in the midplane of the disc, but the explosion velocity is greater than the escape velocity, resulting in a significant portion of the ejecta escaping from the disc. The ejecta may also be partially decelerated before it rushes out of the disc so that the ejecta can fall back. If so, the explosion height needs to be precisely adjusted, which seems very unphysical.

For lower mass SMBHs, the light curve and SED at peak luminosity need to be corrected. As shown in Fig. ~\ref{fig:massVSx}, the fallback ejecta at the inner radius of a SMBH with $M_{\rm SMBH}\sim10^6M_\odot$ is sufficiently massive that it can induce changes in the structure of the AGN disc. Considering that not all the kinetic energy of the fallback ejecta is converted into thermal energy, some part of the fallback ejecta may deviate from a blackbody spectrum. However, it is still a reliable approximation to estimate the SED at this time due to the symmetrical nature of the fallback ejecta around the AGN disc. As a result, the Z-component of the velocity cancels out, and only a small fraction of the kinetic energy is consumed by the AGN disc.

In Section ~\ref{sec:4}, we demonstrate that blackbody emission necessitates an optically thick medium but also requires efficient emission processes. The caveat of this key assumption should be discussed in more details. If the ejecta falls back at a relatively small distance ($\lesssim10^3 r_{\rm g}$), free-free cooling dominates, but insufficient high-energy photons from free-free radiation can cause the spectrum to deviate from the blackbody spectrum. This results in a Wien spectrum at high frequencies dominated by inverse Compton and a flat spectrum at low frequencies dominated by free-free emission \citep{2010ApJ...725..904N, 2023arXiv230316231L}. Nonetheless, it is expected that this deviation to not be significant because there are enough low-energy photons \citep{2010ApJ...716..781K} to couple with hot electrons, and soft photons from the AGN disc can serve as seed photons for the inverse Compton scattering. 
If the fallback position is farther away ($\gtrsim 10^3 r_{\rm g}$), the fallback ejecta become highly dispersed and optically thin, resulting in the domination of inverse Compton radiation, with the seed of soft photons originating mainly from the AGN disc. However, the actual radiation spectrum is complex, and the above-mentioned radiation mechanisms can coexist and evolve with time. Various corrections, such as the generation of positron-electron pairs and possible magnetic effects, should also be considered. Although the temperature calculated from Equations (~\ref{eq:5}) and (~\ref{eq:7}) can reach up to $\sim10^{11}\rm\ K$, we have ignored the pair production. Therefore, the multi-temperature blackbody used in this paper as a preliminary approximation to the complex radiation process has some limitations, and further study of the detailed radiation processes and radiation transfer is necessary to obtain the more realistic spectrum.

\subsection{Potential implications}

 In this study, the calculation of trajectories and the disc model does not incorporate general relativity (GR). It is assumed that collisions occur at distances $\lesssim10^3 r_{\rm g}$; these collisions may have already taken place before the WDs reach the innermost region. For example, at $\sim100r_{\rm g}$, GR effects (the change of fallback time of debris and the change of relative velocity due to the precession) are far less than other effects such as SMBH mass and explosion velocity. Therefore, as a leading-order result, it is reasonable for us to calculate trajectories in Newtonian mechanics. However, GR calculation is indispensable if some collisions occur in the very inner region ($\sim10r_{\rm g}$). Such reasoning should also apply to the AGN disc model; the effect of GR on the disc model is also the sub-leading effect outside the very inner region. In addition, as long as the accretion rate is not too high, the difference between the actual angular velocity of the disc gas and the Keplerian angular velocity is very small \citep{2021MNRAS.505.1324P}, so the model of \citet{kato1998black} is sufficient for our discussions.

 If not all WDs collide outside the very inner region, some WDs can continue on their paths until they collide in the last migration trap \citep{2021MNRAS.505.1324P} or fall into the SMBH. When WDs collide in the very inner region, additional GR effects are expected. One such effect is related to the optical radiation of SN Ia, which mainly originates from radioactive decay. The decay time is measured in the local proper time and, as a result of GR, the decay time seen by distant observers might appear longer compared to that of a normal SN Ia. It should be noted that observing this effect is challenging, since one needs to wait for a certain amount of time after the peak luminosity to measure it, during which the luminosity is already lower than the AGN background.

Furthermore, there are other potential effects to consider. First, an extreme mass ratio inspiral (EMRI) occurs before the collision of the WDs, accompanied by the corresponding GW signal. If the WDs collide in the very inner region, the GW signal abruptly disappears after the collision. Second, the initial explosion can cause significant destruction to the AGN disc, resulting in a substantial decrease in AGN luminosity, while the SN Ia luminosity, particularly in the optical band, increases significantly. Third, in the very inner region of the AGN disc, where the explosion velocity is much lower than the escape velocity, the accretion of fallback ejecta can form an accretion disc. This transient source exhibits similarities to a TDE. Quantitative calculations of these effects would require the application of general relativistic magnetohydrodynamics, coupled with radiation transfer simulations. However, such detailed techniques extend beyond the scope of this paper and will be addressed in future work.

The destruction of the AGN disc can result in an elevated accretion rate and fluctuations in AGN luminosity, potentially leading to the manifestation of a changing-look AGN. The radiation from the initial explosion and its impact on the AGN disc should be considered. The explosion is similar to normal SN Ia, but the AGN's radiation and the subsequent effects should also be taken into account. Based on the result from \citet{2000A&A...359..876C}, the initial explosion luminosity of SNe Ia can reach $\sim10^{43}\rm\ erg\,s^{-1}$. In the example provided by \citet{2021ApJ...906L..11Z}, when the AGN accretion rate is 0.3 times the Eddington accretion rate and the SMBH mass is $\lesssim10^{7}M_{\odot}$, the AGN optical luminosity is $\lesssim10^{43}\rm\ erg\,s^{-1}$. Therefore, within a reasonable range of parameters, the optical luminosity of AGN  can be lower than that of a SN Ia explosion, although some may be comparable. The motion of the radioactively decaying ejecta is dominated by the SMBH gravity, which changes the optical depth relative to the observer \citep{2016ApJ...819....3M}. The interaction between the disc gas and the supernova ejecta soon after the explosion may lead to a fast early peak in the light curve \citep{2016ApJ...826...96P,2021ApJ...923L...8J}. Another concern is whether the initial explosion can damage the AGN disc significantly. How large is the cavity created by the initial explosion? For our fiducial model in which the initial ejecta sweeps from the AGN disc a mass comparable to its own, we estimate that the radius of the cavity is only $\sim 5r_{\rm g}$. Under these circumstances, the AGN disc experiences only slight damage at the position of the event ($\gtrsim100r_{\rm g}$). However, for less massive black holes or thinner discs, the initial explosion may have a significant impact on the disc, and these considerations deserve further investigation.

Our model has broader implications for other explosive events occurring in AGN discs. 
The kinetic energy of the explosive ejecta in an AGN disc plays a critical role, especially in the inner regions. These explosive events may involve collisions between WDs and BHs or NSs, resulting in micro-TDEs \citep{2016ApJ...823..113P,2021arXiv210502342Y}. Furthermore, AGN discs can also host other types of collisions, such as NS-NS and NS-BH collisions, or the formation of binaries with very high eccentricities.  
These events can lead to the ejection of larger amounts of material compared with typical kilonovae \citep{2016MNRAS.460.3255R,2012PhRvD..85l4009E,2014PhRvD..90b4026F,2015PhRvD..92d4028K}, and produce unique GW signatures \citep{2018PhRvD..98j4028P,2019PhRvD.100f3001V,2020PhRvD.102h3005W}. Additionally, AGN discs could potentially serve as sites 
for the production of heavy elements. For instance, the fallback of SN Ia ejecta onto AGN discs significantly enhances the disc's metallicity and may even contribute to the synthesis of heavy elements within the disc.

\section{Conclusions}
\label{sec:6}

In the inner region of an AGN disc, the collision between two WDs can lead to a SN Ia explosion through Jacobi capture. 
This collision is distinct from typical SN Ia events due to the combined effects of the disc gas and the gravitational influence of 
the SMBH.
During the collision, the explosion energy is primarily stored in the ejecta as kinetic energy. Unlike in normal SN Ia events, the AGN disc gas does not decelerate the kinetic energy of the ejecta effectively. Instead, the ejecta rapidly escapes the disc, influenced by the strong gravitational pull of the SMBH. However, a significant portion of the ejecta experiences fallback, returning to the disc. This fallback ejecta is dispersed both in time and spatial position within the disc.

The subsequent interaction between the fallback ejecta and the fast-rotating disc leads to the conversion of kinetic energy into high-energy thermal radiation. This radiation is predominantly emitted in the hard X-ray to soft gamma-ray range. 

	$\bullet$ In the fiducial model with parameters $M_{\rm SMBH}=10^7M_\odot$, $R_{\rm ej}=100r_{\rm g}$, $v_{\rm ej}\sim1.5\times 10^4\rm\ km\,s^{-1}$, the peak luminosity resulting from the fallback ejecta impacting the disc can reach approxiately $\sim10^{45}\rm\ erg\,s^{-1}$. Following the peak, the light curve exhibits a rapid decline with a power-law dependence of $L\propto\,t^{-2.8}$, followed by a relatively gentler decay with $L\propto\,t^{-2.0}$. The highest rate of ejecta mass falling back to the AGN disc occurs at a distance of approximately $\sim 20r_{\rm g}$, and the corresponding fallback time for these ejecta is relatively short, aligning with the peak of the light curve. These events may be related to AGN variations on times-cales of hours to weeks.
	
	$\bullet$ The variability time-scale in the light curve depends on the mass of SMBH. Specifically, the time-scale increases by approximately an order of magnitude when the SMBH mass increases by an order of magnitude. However, the overall power-law shape of the light curve remains unchanged. This suggests that the underlying mechanisms governing the variability are determined primarily by the dynamics of the WD collisions and their interaction with the AGN disc, rather than the specific SMBH mass. In the case of an AGN disc with $M_{\rm SMBH}\lesssim10^7M_\odot$, the fallback ejecta at the inner radius of the disc can be sufficiently massive to destroy the disc, resulting in a changing-look AGN.
    The explosion velocity also plays a role in shaping the light curve. A larger explosion velocity leads to a smaller fallback ratio of the ejecta and an earlier time of peak luminosity. However, the overall shape of the light curve remains unchanged. The higher peak luminosity is attributed to the larger explosion velocity, which results in a larger relative velocity between the fallback ejecta and the disc gas.
    The explosion radius also impacts the light-curve profile. A larger explosion radius leads to a smaller fallback ratio, indicating a reduced mass of ejecta falling back onto the AGN disc. This results in variations in the shape of the light curve. 
    On the other hand, different ejecta masses mainly influence the peak luminosity of the light curve, with no significant impact on the profile of the light curve or the time it takes to reach its peak. The overall shape of the light curve remains consistent, while higher ejecta masses contribute to higher peak luminosities.
	
	$\bullet$ The SEDs of these explosive events, characterized by multi-temperature blackbody spectra that peak in the hard X-ray to soft $\gamma$ range, could be potentially observed by the future $\rm MeV$ detectors. During the maximum light phase, the peak frequency of the SED is higher compared to later times. This is attributed to the larger relative velocity between the fallback ejecta and the gas within the AGN disc. Specifically, for AGN discs with $M_{\rm SMBH}=10^6M_\odot$, the higher peak frequency of the SED arises from collisions between symmetrical fallback ejecta. These collisions contribute to the generation of higher energy photons, resulting in a shift towards higher frequencies in the SED.

Our model has potential implications for other explosive events in AGN discs. By combining future multi-messenger observations (e.g., gravitational waves, neutrinos), we could obtain a more detailed and comprehensive picture of the dynamics and properties of these events, which will provide valuable insights into important topics including AGN variability, the nature of changing-look AGNs, and the origin of heavy elements within AGN environments.

\section*{Acknowledgements}
We thank the referee for critical suggestions and comments, which are very helpful for the improvement of this work.
YFY was supported by the National SKA Program of China
No. 2020SKA0120300, and the National Natural Science Foundation of
China (Grant No. 11725312).
LCH was supported by the National Science Foundation of China (11721303, 11991052, 12011540375, 12233001) and the China Manned Space Project (CMS-CSST-2021-A04, CMS-CSST-2021-A06).
JMW acknowledges support from 
the National Key R\&D Program of China (2020YFC2201400, 2021YFA1600404), NSFC (NSFC-11991050, -11991054, -11833008). 

\section*{Data Availability}
The data underlying this article will be shared on reasonable request to the corresponding author (YFY).



\bibliographystyle{mnras}
\bibliography{BWD-radiation} 



\bsp	
\label{lastpage}
\end{document}